\documentclass[showpacs,twocolumn,aps,preprintnumbers,letterpaper]{revtex4}
\usepackage{amsmath,amssymb}
\usepackage{epsfig}
\usepackage{graphicx}
\usepackage{amsmath}
\usepackage{slashed}
\usepackage{amsfonts}
\usepackage{epstopdf}
\graphicspath{{figures/}}
\usepackage[font=small,labelfont=bf]{caption}

\newcommand{\be}{\begin{equation}}
\newcommand{\ee}{\end{equation}}
\newcommand{\bear}{\begin{eqnarray}}
\newcommand{\eear}{\end{eqnarray}}
\newcommand{\ba}{\begin{array}}
\newcommand{\ea}{\end{array}}

\def\be{\begin{eqnarray}}
\def\ee{\end{eqnarray}}
\def\bea{\be}
\def\eea{\ee}

\def\roughly#1{\mathrel{\raise.3ex\hbox{$#1$\kern-.75em%
\lower1ex\hbox{$\sim$}}}}

\begin{document}

\title{QCD Dirac Spectrum at Finite Chemical Potential:\\
Anomalous Effective Action, Berry Phase and Composite Fermions}

\author{Yizhuang Liu}
\email{yizhuang.liu@stonybrook.edu}
\affiliation{Department of Physics and Astronomy, Stony Brook University, Stony Brook, New York 11794-3800, USA}
\author{Ismail Zahed}
\email{ismail.zahed@stonybrook.edu}
\affiliation{Department of Physics and Astronomy, Stony Brook University, Stony Brook, New York 11794-3800, USA}


\date{\today}
\begin{abstract}
We show that the QCD Dirac spectrum at finite chemical potential using  a matrix model
in the spontaneously broken phase,
is amenable to a generic 2-dimensional  effective action.  
The eigenvalues form a droplet with strong screening and plasmon oscillation. 
The droplet is threaded by a magnetic vortex which is at the origin of a Berry phase. 
For quarks in the complex or Dirac representation, the anomalous transport in the droplet of eigenvalues
bear some similarities with that in droplets of composite fermions at half filling suggesting that the latters  maybe Dirac fermions.
 \end{abstract}






\maketitle

\setcounter{footnote}{0}



%

\section{Introduction}

QCD breaks spontaneously chiral symmetry with a wealth of evidence 
in hadronic processes at low energies~\cite{BOOK}. First principle
lattice simulations strongly support that~\cite{LATTICE}.  The spontaneous
breaking is characterized by a large accumulation of  eigenvalues
of the Dirac operator near zero-virtuality~\cite{CASHER}. 
The zero virtuality regime is ergodic, and its neighborhood is diffusive~\cite{DISORDER}.

The ergodic regime of the QCD Dirac spectrum 
 is amenable to a chiral random matrix model~\cite{SHURYAK}.  In short, the model simplifies
the Dirac spectrum to its zero-mode-zone (ZMZ). The Dirac matrix is composed of hopping between 
N-zero modes and N-anti-zero modes because of chirality, which are  Gaussian sampled by the
maximum entropy principle.  The model was initially  suggested as a null dynamical
limit  of  the random instanton model~\cite{MACRO}.

QCD at finite chemical potential $\mu$ is subtle on the lattice due to the sign problem
\cite{SIGN}.  A number of effective  models have been proposed to describe the effects of matter in QCD with
light quarks~\cite{BOOK}. Chiral random matrix models offer a simple construct 
that retains some essentials of chiral symmetry both in vacuum and matter. For instance, 
in the chiral 1-matrix model finite $\mu$ is captured by a constant deformation of Gaussian matrix 
ensembles~\cite{STEPHANOV,US}. In the chiral 2-matrix model the deformation with $\mu$ is also
random~\cite{OSB,AKE}. Chiral matrix models in matter were discussed by many~\cite{BLUE,RMANY}.
Recently both a universal shock analysis~\cite{NOWAK} and a hydrodynamical description of
the Dirac spectra were suggested~\cite{HYDROUS}.

In the ergodic regime the 1- and 2-matrix models exhibit the same microscopic universality for vanishingly
small $\mu^2$ in the large volume limit~\cite{RMANY}. This limit corresponds to a weak non-hermitean
deformation of the standard chiral matrix models and therefore preserves the underlying chiral symmetry 
of the coset space. It follows the general strictures of the epsilon-expansion in chiral power 
counting~\cite{LEUT}. The microscopic universality for small $\mu^2$ leads to new Leutwyler-Smilga
sum rules~\cite{LS} for the eigenvalues of the QCD Dirac operator that put first principle constraints on the
effective Lagrangian approach in matter as well QCD lattice simulations in the regime of small quark masses, 
small $\mu^2$ and large volumes.

In the first part of the paper, we will show that at finite $\mu$ the distribution of  Dirac eigenvalues in the complex plane 
maps onto a 2-dimensional Coulomb gas whose effective action is constrained  by Coulomb$^\prime$~s law, 
the conformal and gravitational anomalies in 2-dimensions. These  contributions are generic and go beyond
the specifics of matrix models at finite $\mu$. The mapping offers a physical framework  through anomalies,
for understanding first principle 
aspects of the QCD Dirac spectrum at finite $\mu$  with the hope of constraining further the effective
Lagrangian approaches and current and future lattice simulations near the chiral point at finite $\mu$ and for large volumes.
We will also show that  the  2-dimensional Coulomb gas of eigenvalues exhibits both screening  and  a plasmon
branch. The  latter translates to a diffusive mode in the stochastic evolution in chiral matrix models~\cite{HYDROUS}. 
The plasmon frequency sets  an estimate for  the relaxation time for the restoration/breaking of chiral symmetry
at finite $\mu$, solely through the stochastic re-organization of the low-lying  modes of the Dirac spectrum.

In the second part of the paper, we follow by noting  that under adiabatic changes the complex eigenvalues 
at finite $\mu$, viewed as particles in the
complex 2-plane,  behave as fermions for $\beta=2$ due to the emergence of a Berry phase of $\pi$.
There is no Berry phase for $\beta=1$ or when the quarks are in the real representation. We use this observation
to note that a dynamical droplet of quark eigenvalues for $\beta=2$ and finite $\mu$ share some similarities 
with composite fermions at half filling suggesting that the latters are Dirac particles.
We show that the anomalous charged transport contributions from an induced Wess-Zumino-Witten term 
are consistent with constituents of charge $e=1$ and spin $s=1/2$. We use these observations to derive novel effects
for the composite fermions at half filling.

This paper consists of the following new results:
1/ a generic anomalous effective action for the QCD Dirac eigenvalues as a droplet in the complex 2-plane of eigenvalues
with scalar curvature;
2/ a plasmon dispersion law in the bulk of the droplet 
with an improved  estimate for the relaxation time for the breaking/restoration
of chiral symmetry in QCD at finite $\mu$;
3/ an identification of the eigenvalue droplet as a 2-dimensional Fermi liquid threaded by  a magnetic vortex;
4/ a description of the anomalous transport on the Fermi surface through a Berry induced Wess-Zumino-Witten type term;
5/ a suggestion through geometry that composite fermions in the fractional quantum Hall effects at
half filling are Dirac particles; 6/ two novel anomalous transport effects in composite fermions at half filling, caused by a
rotation or a temperature gradient.


%

\section{2-Matrix Model}

The  low lying eigenmodes of the QCD Dirac operator capture some aspects of the spontaneous
breaking of chiral symmetry both in vacuum and in matter.  Remarkably, their fluctuations  follow
by approximating the entries in the Dirac operator by purely random  matrix elements which 
are chiral (paired spectrum) and fixed by time-reversal symmetry (Dyson ensembles).
At finite $\mu$ the Dirac spectrum on the lattice is complex~\cite{BARBOUR}. The matrix
models at finite $\mu$~\cite{STEPHANOV,OSB} capture this aspect of the lattice
spectra and the nature of the chiral phase transition~\cite{BOOK,BLUE,RMANY}. For a 2-matrix model,  
the partition function is~\cite{OSB,AKE}

\begin{eqnarray}
\label{Z2}
&&Z[\beta,\mu, N_f, m_f] =\nonumber\\
&&\int d{\bf A}\,d{\bf B}\, e^{-aN{\rm Tr}( {\bf A}^\dagger {\bf A})}\, e^{-aN{\rm Tr} ({\bf B}^\dagger{\bf  B})}\nonumber\\
&&\times {\rm det}\left( \begin{array}{cc}
-im_f & {\bf A}-i\mu {\bf B} \\
{\bf A}^\dagger-i\mu {\bf B}^\dagger & -im_f
\end{array} \right)^{N_f}
\end{eqnarray}
for equal quark masses $m_f$  in the complex representation. Here ${\bf A}, {\bf B}$ are $C^{(N+\nu)\times N}$ valued.
$\nu$ accounts for the difference between 
the number of zero modes and anti-zero modes. The vacuum Banks-Casher formula~\cite{CASHER} fixes the
dimensionfull parameter $\sqrt{a}=|q^\dagger q|_0/{\bf n}$ in terms of the massless quark condensate  and
the density of zero modes ${\bf n}=N/V_4$. Throughout, we will set the units using $\sqrt{a}\rightarrow 1$. 
All canonical units are recovered  by inspection.

 \begin{figure}[h!]
  \begin{center}
  \includegraphics[width=7cm]{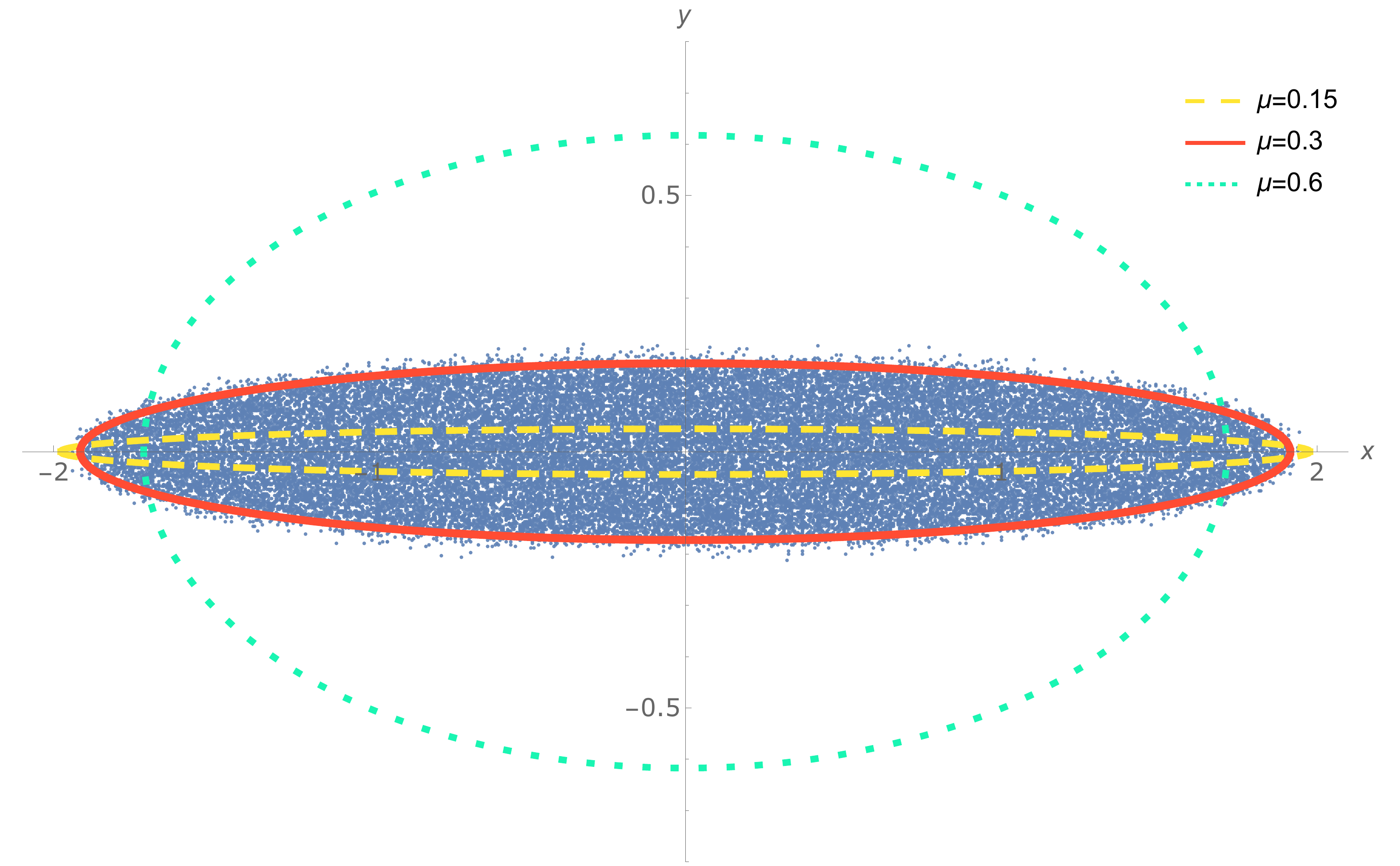}
   \caption{Eigenvalue distribution from a 2-matrix model. }
     \label{fig_density2}
  \end{center}
\end{figure}


The Dirac matrix in (\ref{Z2}) has $\nu$ unpaired zero modes and $N$ 
paired eigenvalues $\pm iz_j$  in the massless limit. The paired eigenvalues
delocalize and are represented by (\ref{Z2}). The unpaired zero-modes decouple.
In terms of the fixed eigenvalues and large but finite $N$, (\ref{Z2}) reads~\cite{OSB,AKE},

\bea
\label{1}
Z[\beta, \mu,  N_f, m_f]\approx &&
\int \prod_{i=1}^Nd^2z_i\,\left(z_i^2+m_f^2\right)^{N_f}\,e^{-S[\beta, \mu; z]}\nonumber\\
\eea
The action is

\bea
\label{2}
S[\beta, \mu; z]=-\beta\sum^N_{i<j=1}{\rm ln}|z^2_i-z^2_j|
+\sum_{i=1}^NW(z_i)
\eea
with 

\bea
\label{3}
W(z)=-\alpha\,{\rm ln}|z|+ \frac{B}2w(z)
\eea
and the quasi-harmonic potential $w(z)=|z|^2-\frac{\tau}2 (z^2+{\overline z}^2)$.
Here $\alpha=\beta(\xi+1)-1$ and $\xi$ accounts for the difference between 
the number of zero modes and anti-zero modes in the ZMZ. We define 

\be
\frac B{N\beta}=\frac 1{1-\tau^2}=\frac{1+\mu^2}{2\mu^2}\equiv \frac 1{l^2}
\label{BM}
\ee
with $B=1/l_B^2$ acting  as a magnetic field with magnetic length $l_B$
as we will suggest  below.
Throughout  $\beta=2$  unless indicated otherwise.
The 2-matrix model for  $\beta=1,4$~\cite{WAY} is 
more subtle at finite $\mu$~\cite{AKE}.

In Fig.~\ref{fig_density2} we display the distribution of eigenvalues following from the 2-matrix model 
with ${\bf A}$ and ${\bf B}$ sampled from a Gaussian ensemble of $200\times 200$ matrices with $\nu=0$ and 
$\mu=0.3$.  The mean density in the droplet is ${\rho}_0\approx \nu B/2\pi$ with $\nu=1/\beta$.
The  boundary curves follow from the analysis in~\cite{BLUE}. 
The domain is an ellipse 

\be
\label{ELL}
\frac {x^2}{a_+^2}+\frac{y^2}{a_-^2}=1
\ee
with semi-axes ${a_\pm^2}/{2l^2}={1\pm \tau}/{1\mp \tau}$.
The ellipse remains  un-split with area  ${\cal A}=\pi a_+a_-=2\pi l^2$ for all values of $\mu$.
For the other quark representations with $\beta=1,4$ the joint distribution
in the 2-matrix model is more subtle~\cite{AKE}. Throughout, (\ref{1}) will be assumed
for $\beta=2$,  but most results  extend to $\beta=1,2,4$ for large $N$.

For completeness we recall that the 1-matrix model corresponds to setting ${\bf B}={\bf 1}$. 
In this case, the eigenvalue distribution forms
a connected droplet in the z-plane for $\mu<\mu_c$, and splits into two  
droplets symmetric about the real-axis for $\mu>\mu_c$,  restoring chiral symmetry~\cite{STEPHANOV,US}. 
In the spontaneously broken phase, all droplets are connected and symmetric about the
real-axis


Throughout,  (\ref{1}) is to be understood in the large $N$ limit to allow for a course
graining of the eigenvalue density. We note that in this limit,  the Bessel kernel
in~\cite{OSB} is expanded.  While the 2-matrix model is exactly solvable in terms 
of the orthogonal polynomial method, our analysis of this model shows the emergence
of a generic effective action for the complex eigenvalues of the Dirac spectrum at
finite $\mu$. Therefore our analysis encompasses the weak non-hermiticity range of the model.
Besides chiral symmetry through the pairing of the complex eigenvalues, 
the  new guiding principles for the construct of
this effective action are Coulomb law, the conformal and gravitational anomalies
in 2-dimensions as we now detail.



\section{ Static Effective Action}

In this section we will re-write (\ref{1}) in terms of the effective  potential generated by the
mean charged density, for a sufficiently dense ensemble of eigenvalues. This
assumes that $N$ is large to allow for a smoothening of the eigenvalue density over
distances larger than the inter-level spacing but much shorter than the size of the
eigenvalue droplet. While the 2-matrix model is solvable~\cite{AKE,OSB}, it is
important to stress that its re-writing at large but fixed $N$, using an effective action 
unravels universal physical aspects of the eigenvalue droplet that are not restricted 
to the specifics of  the model. We will be able to go beyond the strictures of
chiral symmetry by making use of emergent anomalies when using the collective
potential sourced by the smoothened eigenvalue density as we detail below.

For the sake of generality, we will assume that the eigenvalue space is curved
with a measure $ds^2=g_{z\bar z}dzd\bar z$.  The explicit form of the metric is not
important for our general arguments. The curved eigenvalue space will allow for 
the unraveling  of two generic contributions to the effective action through the 
emergence of conformal and gravitational anomalies, that make the effective
action construct more general than the 2-matrix model.
Also, for bulk quantities a fixed eigenvalue curvature $R$ is conjugate to 
a fixed space curvature $R_4\approx 1/R$.

 The re-writing of (\ref{1}) will closely follow the effective action construction 
 in~\cite{CAN} for the Lauhglin states to which we refer for further details.  In brief, 
on a curved  eigenvalue  manifold of volume $V_2$ with a metric $g_{z\bar z}$ and
large $N$, the ensemble described by (\ref{3}) allows the change in the measure,
from integrating over specific eigenvalues $z_i$ to integrating over the eigenvalue 
density $\rho(z)$. Specifically

\be
\prod^N_{i=1}\sqrt{g}\,dz_i\rightarrow e^{\int  dz\sqrt{g}\rho(z){\rm ln}(\rho_0/\rho(z))} D\rho 
\label{11}
\ee
with the induced Boltzmann entropy in the exponent~\cite{MEHTA,CAN}.
The mean density $\rho_0$ will be made explicit below.
Since the spectrum is chirally symmetric, we reset 
$(\rho(z)+\rho(-z))/2\rightarrow \rho(z)$ for convenience. 
We will assume that the density and thus the potential are real,
which effectively corresponds to a phase quenched approximation.
This approximation while limited in bulk~\cite{STEPHANOV,US},
still contains useful physical information at the edge of the
Dirac spectrum~\cite{LIU4}.

Following~\cite{CAN},
we substitute the collective potential $\varphi(z)$ to  the collective density $\rho(z)$ through
the Poisson equation

\be
\nabla^2 \varphi (z)= -4\pi\left(\rho(z)-\frac N{V_2}\right)
\label{F1}
\ee
This change of variable involves the conformal anomaly in 2-dimensional
curved space~\cite{CAN,POL}

\be
D\rho\rightarrow e^{-\Gamma_2}\,{\rm det}(-\nabla^2)\,D\varphi
\label{F2}
\ee
with the Liouville action

\be
\Gamma_2=\frac{1}{24\pi}\int dz\sqrt{g} \left((\nabla {\rm ln} \sqrt\rho)^2+R (z)\,{\rm ln}\sqrt\rho\right)\nonumber\\
\label{F3}
\ee
as we briefly detail in the Appendix.
Here $R(z)$ is the scalar  Ricci curvature on the curved 2-dimensional complex manifold
of eigenvalues. The determinant in (\ref{F2}) induces a gravitational anomaly~\cite{CAN}.
The final result for the effective action is

\bea
&&Z[\beta, \mu, N_f, m_f]\approx \left({\rm det}(-\nabla^2)\right)^{\frac 12}\nonumber\\
&&\times \int D\varphi \,e^{-\frac{N_f}{2}\left(\varphi (im_f)+\varphi (-im_f)\right)}\,e^{-\Gamma[\beta; \varphi]}
\label{12}
\eea
with the effective quenched action $\Gamma=E_0+\Gamma_0+\Gamma_1+\Gamma_2$ and

\bea
\label{13}
&&\Gamma_0=\frac{1}{8\pi\nu}\int dz \sqrt{g}\left((\nabla \varphi)^2-R(z) \varphi-4\nu B(z)\varphi\right)\nonumber\\
&&\Gamma_1=\frac{1}{\nu}\left(\nu-\frac{1}{2}\right)\int dz \sqrt{g}\rho \,{\rm ln}\,\rho \nonumber\\
&&E_0=\frac{N}{\nu V_2}\int dz dz^\prime\sqrt{gg^\prime}\,{\rm ln}\,|z-z'|^2\left(\rho_0(z)-\frac 12 \frac{N}{V_2}\right)\, \nonumber\\
\eea
Here $B(z)=B-\pi\alpha\delta(z)$ and $\Gamma_1=0$ for Dirac quarks with $\nu=1/2$.
(\ref{12}-\ref{13}) differs from the one in~\cite{CAN}
in three ways: 1/ Both $\rho(z)$ and $\varphi(z)$ are z-even because of chiral symmetry; 
2/ $B$ is of order $N$; 3/ $B(z)$ carries a magnetic vortex  which will be exploited below. 
We note that $\Gamma$ is real and of order $N^2$.

Although (\ref{12}-\ref{13}) was derived using the 2-matrix model,  we observe that each 
of its contributions are generic and therefore not specific to the 2-matrix model. The 
exception is $w(z)$ which is model specific and in this case quasi-harmonic. 
This suggests that (\ref{12}) is the effective partition function 
for  QCD Dirac spectra at finite $\mu$ in the spontaneously broken phase 
provided that $w(z)$ in (\ref{3}) is extended to include non-quasi-harmonic potentials.


\section{Ground State}

In this section we analyze the ground state properties following from 
(\ref{1}) in terms of the collective potential in (\ref{12}). We will explicit the solution
to $\delta\Gamma/\delta\varphi=0$ in the linearized approximation. The result is an
expression for the mean charged density in the droplet $\rho_0(z)$ without the 
contribution from the $N_f$-exponent in (\ref{12}). We will refer to this solution as the quenched
saddle point which is not to be confused with the standard quenched approximation
using in the litterature.  We will use this result to analyze the mean contribution of the 
low-lying quark eigenmodes in the ZMZ to both the bulk energy and quark condensate
at finite chemical potential and for fixed curvature $R\approx 1/R_4$.
 Again, because of the generic nature of (\ref{12}) as we noted earlier, 
we expect the results to reflect on the QCD eigenvalue spectrum at finite $\mu$ 
beyond the 2-matrix model.

\subsection{Leading}

With this in mind, the  quenched saddle point equation $\delta\Gamma/\delta\varphi=0$ 
in the linearized approximation yields the mean density

\bea
\label{RHO1}
\rho_0(z)\approx \frac{\nu B}{2\pi}+\frac{R}{8\pi}-\frac{\nu \alpha}2\delta(z)\equiv \rho_0-\frac{\nu \alpha}2\delta(z)
\eea
For vanishingly small $\mu^2$, (\ref{RHO1}) yields $\pi \rho_0(z)\approx N/4\mu^2$  in agreement with 
the quenched asymptotic density in the weak non-hermiticity limit~\cite{VERB}. 
The quenched energy in the ZMZ is

\bea
\label{E1}
{\cal E}_0\approx -{\rm ln}{Z}[\beta,\mu, 0, 0]\approx -\frac{1}{2}\int_{\cal A}\sqrt{g} \,\rho_0(z) W(z)
\eea
with pertinent changes in $W(z)$ in curved space.
Using (\ref{RHO1}) we obtain

\bea
{\cal E}_0\approx -\frac{1}{8\pi}\left(\nu {B^2}+\frac 14 {BR}\right)
\int_{\cal A}\sqrt{g}\,w(z)
\label{E0}
\eea
We note the  quadratic form of the magnetic-like contribution,  and the
mixed curvature-magnetic contribution which is a Casimir effect. 
Recently, a similar mixed term between the gauge holonomy and the curvature
in hyperbolic  space was noted in the cosmological context~\cite{ARIEL}. 

The quenched quark condensate in the ZMZ is  ($m_f=0$)

\bea
\label{Q1}
\int_4\left<\bar q  q\right>_0\approx
\left( \frac{\nu B}{2\pi}+\frac{R}{8\pi}\right) 
\int_{\cal A}{\sqrt{g}}\left(i2\pi \nabla_{z^\prime}{\cal G}(z,z^\prime)\right)_{z^\prime=0}
\eea
with  $\nabla^2{\cal G}(z,z^\prime)=\delta(z-z^\prime)/\sqrt{g}$
in curved eigenvalue space. In zeroth order in the curvature 
$2\pi{\cal G}(z,z^\prime)={\rm ln}|z-z^\prime|$ and (\ref{Q1}) vanishes

\be
 V_4\left<\bar q  q\right>_0\approx \frac i2 \left( {\nu B}+\frac{R}{4}\right)  \int_{\cal A}\frac 1z =0
 \ee
 as the last integral is zero when carried over the elliptic droplet (\ref{ELL}),

 \be
 \int_{\cal A}\frac 1z= \int_{-a_{+}}^{a_{+}} dx\, 2\, {\rm sign} (x)\,{\rm tan}^{-1}\left( \frac{a_{-}(x)}{|x|}\right)=0\nonumber
\ee
with  $a_-(x)=a_-\left(1-x^2/a_+^2\right)^{1/2}$.
This result is conform with the  phase quenched limit, with the inside of the
droplet breaking conformal symmetry with a condensation of mixed pairs made of
a quark and a conjugate quark~\cite{STEPHANOV, US}.

\subsection{Sub-leading}

The unquenched but subleading results follow from the linearized saddle point in  (\ref{12}) including the $N_f$-contribution.
Specifically

\be
\label{E2}
\frac{\delta\Gamma}{\delta \varphi (z)}=-N_f\left(1+\frac{N_f\nu}{16\pi{\rho}(z)}\nabla^2\right)
\,|z|\delta(z^2+m_f^2)
\ee
The $N_f^2$  contribution  follows from the arguments presented in~\cite{CAN} for the 
emergence of a conformal dimension associated to $e^{-N_F\varphi/2}$ in (\ref{12}). 
Solving for $\varphi$ in (\ref{E2}) yields the unquenched  density to order ${\cal O}(N_fN^0)$

\be
\label{E3}
\rho_f(z)\approx \rho_0(z)-\frac{\nu N_f}{2}\left(\delta(z+im_f)+\delta(z-im_f)\right)
\ee
The unquenched energy in the ZMZ in flat eigenvalue space is

\bea
\label{E4}
{\cal E}_f
\approx  -\frac{1}{2}\int_{\cal A}\,\rho_f (z)\left( W(z)-N_f{\rm ln}(z^2+m_f^2)\right)
\eea
which differs from  (\ref{E1}) to order ${\cal O}(N_fN)$ by

\bea
\label{E3F}
{\cal E}_f-{\cal E}_0\approx \frac{\nu N_f}4\left(
\frac{ B}{\pi}\int_{\cal A}{\rm ln}(z^2+m_f^2) +\sum_\pm W(\pm im_f)\right)\nonumber\\
\eea
The  unquenched chiral condensate to the same order follows 
from $V_4\left<\bar q q\right>_f=\partial{\cal  E}_f/\partial m_f$. We note that 
$|\left<\bar q q\right>_f| < |\left<\bar q q\right>_0|$. Indeed, (\ref{E3F}) 
yields to order $N_fN$

\be
V_4\left(\left<\bar q  q\right>_f-\left<\bar q  q\right>_0\right)\rightarrow 
{ N_f m_f} \,\frac{\nu B}4\, \left(\frac{a_{+}\pi}{a_{+}+a_{-}}+(1+\tau)\right)
\ee
The effect of the fermion determinant on 
the 2-dimensional droplet amounts to inserting
two static charges of $\nu N_f/2$ at the mirror locations $z=\pm im_f$. The charges 
are strongly screened as we now detail.




\section{ Screening and Plasmons}

In this section we will explicit the screening nature of the droplet of eigenvalues viewed as a 2-dimensional
1-specie plasma with unit charges $e=1$. This is one of the universal feature of the effective action (\ref{12})
which goes beyond the 2-matrix model used. In particular, we will derive the static structure factor of the droplet
using a small and longitudinal deformation of the droplet density. We will observe that the static pole structure
emerging from the structure factor nicely extends time-like to the plasmon pole contribution we have derived
recently using a hydrodynamical analysis~\cite{LIUPIOTR}, with an improved estimate for a relaxation time 
in QCD at finite $\mu$.

\subsection{Static Structure Factor}

\begin{figure}[htb]
\begin{center}
\includegraphics[width=7cm]{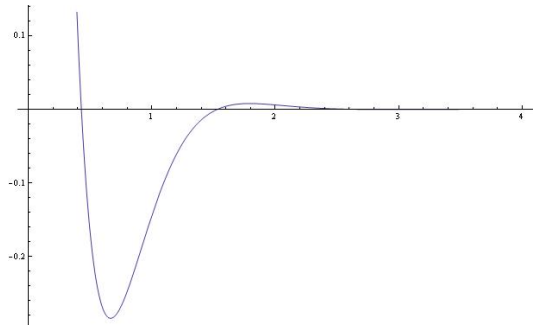}
\caption{Radial structure factor ${\bf G}(r)$ versus $r$.}
\label{GX}
\end{center}
\end{figure}

By rescaling $\tilde{z}=\sqrt N z$ in the microscopic limit, both $B$ and $\rho_0$ are of order 1 and the droplet 
fills out the entire 2-plane in flat space.  Within the center of the droplet, 
the $\varphi (\pm im_f)$-insertions do not affect the density and its response. The electro-static
properties of the droplet are captured by the structure factor

\bea
\label{PLA1}
S(k)=\frac 1N \left<\left|\int dz\,e^{ikz}\,\rho(z)\right|^2\right>_{\rm conn.}
\eea
The longitudinal deformation $\delta \rho(z)\approx -\rho_0 \vec\nabla\cdot\vec\phi$
yields $S(k)\approx {\rho^2_0}{k}^2\left<|\vec\phi(k)|^2\right>$. It follows by
expanding $\Gamma$ in (\ref{12}) to quadratic order in $\delta\rho$. The result is


\be
S(\tilde k)\approx \frac{{\tilde k}^2\omega_p^2}{\omega_p-\frac{\beta -2}4{\tilde k}^2+\frac{\beta}{48}\frac{{\tilde k}^4}{\omega_p}}
\label{PLA2}
\ee
with $\omega_p=B/N$ and the rescaled momentum $\tilde k=k/\sqrt{N}$. 
For $\beta=2$,  the screening length follows from $\tilde{k}_S^4+24\omega_p^2=0$
or $l_S\approx l/\sqrt{N}$. Back to z-space

\be
S(\tilde z)=\int\frac{d^2\tilde k}{4\pi^2}e^{-i\tilde k\cdot \tilde z}S(\tilde k)
=\frac{3\omega_p^3}{\pi}{\bf G}\left(\frac{3|\tilde z|^4\omega_p^2}{32}\right)
\ee
${\bf G}(r)$ is Meijer G-function shown in Fig.~\ref{GX}, with a logarithmic core
${\bf G}(r\ll l_S)\approx -{\rm ln}r$, a hole of range $l_S$ and an  asymptotic  
tail ${\bf G}(r\gg l_S)\approx e^{-2\sqrt{2}r}$.

\subsection{Relaxation Time}

The pole in (\ref{PLA2}) is the static limit of the longitudinal plasmon mode in the droplet.
Indeed, (\ref{PLA2}) together with the hydrodynamical plasmon analysis in~\cite{LIUPIOTR} implies the 
non-linear plasmon dispersion relation for the time-dependent longitudinal modes $\vec\phi(z)\rightarrow\vec\phi(t,z)$

\bea
\label{DX1}
\left(\partial^2_t 
+\left({\omega_p} +\frac{\beta-2}{4}\nabla^2+\frac{\beta}{48}\frac{\nabla^4}{\omega_p}\right)^2\right)\,
\vec{\phi}(t, z )\,\approx 0
\label{PLA3}
\eea
which confirms and extends the hydrodynamical result~\cite{LIUPIOTR}

\bea
\label{DX1X}
\left(\partial^2_t 
+\left({\omega_p} +\frac{\beta-2}{4}\nabla^2
\right)^2\right)\,
\vec{\phi}(t, z )\,\approx 0
\label{PLA3X}
\eea

Now consider a time-dependent deviation  of the eigenvalue density through
$\delta \rho(t,z)\approx -\rho_0 \vec\nabla\cdot\vec\phi (t,z)$ in the
complex 2-plane. For large droplets, the hydrodynamical arguments presented in
~\cite{LIUPIOTR} show that the deformation relaxes through Euler
equations with a large time asymptotics controlled by  the plasmon branch 
(\ref{PLA3X}), i.e. $\delta \rho(t\rightarrow\infty,z)/\rho_0\approx e^{-2\omega_p t}$,
with a relaxation time $T_R\approx 1/2\omega_p$.
If the initial condition for the eigenvalue distribution $\rho(0,z)$ 
is chosen to describe a chirally symmetric phase at finite $\mu$, then the time
it takes for the distribution to relax to the spontaneously broken phase 
is again given by the plasmon branch.
From (\ref{PLA3}) it follows that the relaxation time  for spontaneously breaking/restoring chiral symmetry 
in QCD at finite chemical potential in droplets of finite sizes ${\cal A}=2\pi l^2$ is now of order

\be
\label{TR1}
T_R\approx \frac 1{2\left({\omega_p} -\frac{\beta-2}{4}\frac1{\cal A}+\frac{\beta}{48}\frac{1}{\omega_p{\cal A}^2}\right)}
\ee
with $\nabla^2\approx -1/{\cal A}$ and $\omega_p=B/N=2\pi\beta/{\cal A}$. Although
(\ref{DX1},\ref{TR1})  follow from the model with $\beta=2$, they nicely agree and extend the results
in~\cite{LIUPIOTR} following from the hydrodynamical arguments for the three  Dyson ensembles $\beta=1,2,4$.



\section{Berry Phase}

In this section we will show that if the Dirac eigenvalues where to change adiabatically with
some mathematical  time, i.e. $z_i\rightarrow z_i(t)$, then the effective action will develop among other
thinghs a geometrical contribution of the Wess-Zumino-Witten type. The origin of this term
will be traced to a particular contribution in the eigenvalue measure which distinguishes 
between real ($\beta=1$) or complex ($\beta=2$)  representations for the quarks, i.e. 
whether the underlying quarks are Majorana or Dirac particles. This suggests that an
adiabatically deformed droplet of eigenvalues at finite $\mu$ maps onto a 2-dimensional
fermionic system in a magnetic field.


With this in mind, the magnetic field induced by the 2-matrix model is
$B(z)=B-\pi\alpha\delta(z)$. The first contribution defines the mean
density ${\nu B}/{2\pi}$ in flat space. 
The second contribution is a magnetic vortex  of strength $\alpha/2$ 
($z\rightarrow\vec{z}$)

\be
\label{POLE}
{A}_i^\star=\frac{\alpha}2 \,\frac{\epsilon_{ij}z_j}{{{|z|}}^2}
\ee
(\ref{POLE})  is multivalued and generates a Berry 
phase~\cite{SON1,ZAHED}.
For that, consider an adiabatic time-dependent change in the eigenvalue through 
$\vec z\rightarrow \vec z(t)$ and $\vec v(t)=\vec{z}(t)/|z(t)|$. In
the presence of (\ref{POLE}), an
anomalous Berry contribution is generated for each eigenvalue change as

\be
S_1\equiv \int_1\,{\vec A}^\star\cdot {\dot{\vec z}}\,dt=\frac{\alpha}2 \int_1 \frac{zdz}{|z|^2}=\frac{\alpha}2 \int_1 {vdv}
\label{1BERRY}
\ee
The  form notation is subsumed.  The line integral is over the mathematical time which
is understood as a relaxation time for Dirac spectra in $1+4$-dimensions~\cite{LIUPIOTR}. 
Each of the eigenvalue when adiabatically rotated around the origin 
accumulates a phase $S_1=\alpha\pi$. For 
Dirac quarks in the complex representation and particle-anti-particle symmetry,
$\beta=2$ and $\alpha=1$. The phase accumulation is $\pi$.

To generalize (\ref{1BERRY}) to all particles in the droplet we borrow from  the arguments in~\cite{ZAHED},
and  covariantize and localize $v^i(t)$ in (\ref{1BERRY}) by using the embedding  
$v^i(t)\rightarrow v^\mu(t,z)\equiv (1,\vec v (t,z))$ in $1+2$ dimensions. Thus

\be
S_B=
\frac 12 {\alpha} \rho_0{\cal A}\,\int_1 {vdv}\rightarrow \frac 12 {\alpha} \rho_0\,\int_{1+2} vdv
\label{NBERRY}
\ee
which is a Wess-Zumino-Witten type term.



\section{Fermi Surface}

The complex eigenvalues form a droplet
in the complex 2-plane  for all three Dyson ensembles. However, for $\beta=2$
with the underlying quarks in the Dirac representation, adiabatically changing
eigenvalues behave as moving fermions in 2-dimensions with a fixed magnetic
field $B$ as defined in (\ref{BM}) with filling fraction $\nu=1/\beta=1/2$ since 
the mean droplet density is $\rho_0=\nu B/2\pi$. Note that for $\beta=1$  with  
the underlying quarks in the
real or Majorana representation, the Berry phase is zero with
no identification to fermions. We now explore the consequences of this relationship to 
fermions in 2-dimensions. 


\subsection{Anomalous Fermi Surface}

We identify 
the Fermi momentum  of the droplet as $\rho_0\rightarrow k_F^2/4\pi$ or $k_F\equiv \sqrt{2\nu B}$
for $\beta=2$. The anomalous transport 
on the Fermi surface follows by setting $p^\mu=k_Fv^\mu\rightarrow k_F(v_F,\vec{v}(t,z))$
in (\ref{NBERRY}) and gauging by minimal substitution. The Fermi velocity $v_F<1$. Thus

\be
S_B\equiv \frac {\alpha}{8\pi}\int_{1+2} \sqrt{g}\,(p+eA+sw)\,d(p+eA+sw)
\label{W1}
\ee
Here $A$ is a U(1) gauge field with $F=dA$ and $\omega$ a U(1) 
spin connection with $R=d\omega$.  Similar anomalous effective actions for the quantum Hall effects
were recently discussed in~\cite{CAN,MIXED}. 

The U(1) current $J=\delta S_B/\delta A$ stemming from (\ref{W1}) is anomalous

\be
\frac 1{\sqrt{g}}\, d\,\sqrt{g} \left(J-\frac{\alpha e}{8\pi}\,k_Fv\right)=\frac{\alpha e^2}{8\pi}\,F+\frac{\alpha es}{8\pi}\,R
\label{W3}
\ee
The $k_F=\sqrt{2\nu B}$ contribution in (\ref{W3}) is the analogue of the chiral vortical effect 
in Fermi surfaces threaded by a Berry phase in higher dimensions ~\cite{ZAHED}.
The anomalous contribution to the  density is

\be
\rho_B(z)=\frac 1e\frac{\delta S_B}{\delta A_0(z)}
=\frac{\alpha e}{4\pi} B+\frac{\alpha s}{4\pi} R
\label{W4}
\ee
 A comparison of (\ref{W4}) with $\rho_0$ in (\ref{RHO1}) suggests  that 
 $e=2/\alpha\beta$ and $s=1/2\alpha$. For $\beta=2$ and $\alpha=1$,
 we have $e=1$ and $s=1/2$.  The droplet of eigenvalues maps onto 
 a fermionic droplets in 2-dimensions with constituents of charge $e=1$
 and spin $s=1/2$.
 


\subsection{Relation to Composite Fermions}

 Recently, Son
has argued that in a non-zero magnetic field the composite fermions 
of the fractional quantum Hall effect at half filling,
have a finite density and live in a zero magnetic field~\cite{SON1}. 
He argued that they exhibit particle-hole symmetry and 
that they are Dirac particles. Their ground state is a Fermi liquid 
 with  a  Berry phase of $\pi$. 

 Our construction suggests that the anomalous transport of
 composite fermions in the fractional quantum Hall effect at half filling 
 in $1+2$ dimensions with a magnetic field,  share similarities with  a 2-matrix model of the ZMZ zone of
 Dirac quarks with $\beta=2$ in $1+4$ dimensions in the spontaneously broken 
 phase at finite chemical potential $\mu$. 
The suggestion that composite fermions at half filling behave as Dirac particles
with a Wess-Zumino-Witten term of the type (\ref{W1}) is falsifiable 
as it leads to specific and measurable predictions as we now detail.

\subsection{Rotating Fermi Surface}

We recall that novel chiral vortical effects were noted  in rotating Weyl Fermi liquids
with a Berry phase in $1+3$ dimensions~\cite{ZAHED}.  
We now show that similar effects are expected for composite
fermions  at  filling fraction $\nu=1/2$ in $1+2$ dimensions if they are Dirac
particles. Throughout this sub-section $B$ is understood as a real (residual) magnetic field.

Consider now that the 2-dimensional fermions are in 
a rotating frame along the z-direction 
with velocity $\vec{\Omega}=\Omega\hat{z}$. Each
fermion experiences centrifugation that is best 
captured  by the  gravito-electro-magnetic fields
$E_g=p^0\vec\nabla \theta$ and $\vec B_g=p^0\vec \Omega$
with a metric $g_{00}=1-\Omega^2|z|^2+2\theta$~\cite{LANDAU}. 
$\theta$ acts as a gravitational-like  potential. In the Fermi sea, the 
inertial 2-force on  a quasiparticle of 3-momentum $p^\mu\rightarrow  (\epsilon, s\vec {k}_F)$ 
is the Lorentz-like force~\cite{ZAHED}

\be
\vec F_g=\epsilon \vec\nabla \theta +\vec {k}_F\times {\bf s}\vec\Omega
\ee
with the Coriolis force manifest. Here ${\bf s}=\pm$ for a particle  or anti-particle (hole).
The corresponding Berry induced mixed Chern-Simons term in the Fermi liquid droplet is
constructed following the arguments in~\cite{ZAHED} to which we refer to for more details. 
Since $e\alpha=2/\beta=2\nu$, the result in our case is 

\bea
\label{OMEGAZ}
S_\Omega=\frac{\nu}{4\pi}\left(\sum_{{\bf s}=\pm }\int_0^\infty\frac{d({\bf s}\epsilon)}\epsilon\, \epsilon\,{\bf f}(\epsilon, {\bf s})\right)
\nonumber\\
\times 2\int_{1+2}\,\left(A_0\,{\bf s}\vec{ \Omega}+(\vec A\times\vec\nabla\theta)\right)_z
\eea
The Fermi distribution ${\bf f}(\epsilon, s)=1/(1+e^{({\epsilon-{\bf s}\mu_F)/T})}$ 
for the composite fermions with particle-hole symmetry satisfies

\bea
&&\left(\sum_{{\bf s}=\pm }\int_0^\infty{d (s\epsilon)}\,{\bf f}(\epsilon, {\bf s})\right)=\mu_F=k_F=\sqrt{2\nu B}\\
&&\left(\sum_{{\bf s}=\pm }\int_0^\infty{d (\epsilon)}\,{\bf f}(\epsilon, {\bf s})\right)=T\,{\rm ln}\left(2+2\,{\rm ch}\left(\frac{k_F}T\right)\right)\nonumber
\ee

(\ref{OMEGAZ}) yields an anomalous and $\Omega$-driven contribution to the 
composite fermion  density

\bea
\label{ROMEGAZ}
\rho_\Omega(z)=\frac{\delta S_\Omega}{\delta A_0}=
\frac{\nu}{2\pi}\,T\,{\rm ln}\left(2+2\,{\rm ch}\left(\frac{k_F}T\right)\right)\,\Omega
\eea
essentially from  the Coriolis force contribution. The limiting results are measurable shifts in the 
density

\bea
\label{ZRHOZ}
&&\rho_\Omega(z)\rightarrow \frac{\nu T{\rm ln}2}{\pi}\Omega\qquad  T\gg k_F=\sqrt{2\nu B}\nonumber\\
&&\rho_\Omega(z)\rightarrow \frac{\nu \sqrt{2\nu B}}{4\pi}\Omega\,\,\,\,\,\,\,\,T\ll k_F=\sqrt{2\nu B}
\eea

The $\theta$-contribution in (\ref{OMEGAZ})
can be exploited using an observation by Luttinger who noted that the effect
of a temperature gradient can be balanced by a gravitational potential~\cite{LUTTINGER}. The response
of the composite fermions in a Fermi disc to a small temperature gradient can be captured by
$\vec\nabla\theta=-\vec\nabla{\rm ln}T$. As a result, an anomalous and measurable stationary  gradient flow develops 

\be
\label{JOMEGAZ}
J_{\Omega,i}(z)=\frac{\delta S_\Omega}{\delta A_i}=\frac{\nu \sqrt{2\nu B}}{2\pi}\,\epsilon_{ij}\nabla_j{\rm ln}\,T
\ee

The results of this  sub-section rely only on the interpretation that composite
fermions are Dirac particles with particle-antiparticle (hole) symmetry and 
the origin of the Wess-Zumino-Witten term (\ref{W1}), with $B$ a real (residual)
magnetic field as we noted earlier. Therefore they are more general than the correspondence we have developed.

\section{Conclusions}

The QCD Dirac spectrum at finite chemical potential 
following from a 2-matrix model at large $N$ maps onto a generic but anomalous effective theory 
in 2-dimensions with both gauge and conformal
anomalies in bulk, much like in the fractional quantum Hall effect~\cite{CAN}. 
The complex eigenvalues form a droplet  threaded by 
a magnetic vortex  at the center. Adiabatic changes in the Dirac eigenvalues 
are characterized by a Berry phase that leads to a Wess-Zumino-Witten term.
The anomalous transport of composite fermions 
at half filling in $1+2$ dimensions share similarities with a 2-matrix model of Dirac particles in
$1+4$ dimensions in the spontaneously broken phase at finite chemical potential.


\section{\ Acknowledgements}

We thank Sasha Abanov and Piotr Warchol for discussions.
This work  is  supported in part  by the U.S. Department of Energy under Contracts No.
DE-FG-88ER40388.

\section{Appendix: Conformal Anomaly}

The emergence of the Liouville effective action in (\ref{F2}-\ref{F3}) follows from the conformal
anomaly~\cite{CAN,POL}. In this Appendix, we give a brief account of the derivation following
~\cite{CAN}. The change from the collective density $\rho$ to the collective potential $\varphi$
follows formally from (\ref{F1}) as

\be
\label{A1X}
{\rm det}(-\nabla^2)\,D\varphi=\left(\int D\eta D\bar\eta \,e^{\int dV_2\, \bar\eta\,\nabla^2\, \eta}\right)\,D\varphi
\ee
with $\eta, \bar\eta$ as Grassmanians.
The regularized measure $D\eta D\bar\eta D\varphi$ depends implicitly on the collective density $\rho$
and thus the collective potential $\varphi$ through regularization~\cite{CAN,POL}. 
As a result, $\rho$ can be treated as a conformal factor of the 
mathematical metric $g_{z\bar z}$ on the eigenvalues, and removed from the measure by a conformal transformation
of the coordinates 

\be
dV_2\equiv \sqrt{g}\,dz\rightarrow dV_2/\rho
\ee
The result is (\ref{F2}-\ref{F3}) after using the 
central charges $c_\varphi=+1$ and $c_\eta=c_{\bar\eta}=-1$.

 \vfil
\end{document}